\documentclass[twocolumn,11pt,a4paper]{article}
\usepackage{latexsym}
\usepackage{times}
\usepackage{helvet}
\usepackage{courier}
\usepackage{amsmath}
\usepackage{mathptmx}
\usepackage{amssymb}
\usepackage[utf8]{inputenc}
\usepackage{xcolor}
\usepackage{color}
\usepackage{url} 
\usepackage{multirow,graphicx}

 \usepackage{float}
 \usepackage{footnote}
\usepackage{listings}

\long\def\NBB#1{}

\def\lpmln{{\rm LP}^{\rm{MLN}}}

\long\def\BOC#1\EOC{\message{(Commented text )}}
\long\def\BOCC#1\EOCC{\message{(Commented text )}}
\long\def\BOCCC#1\EOCCC{\message{(Commented text )}}
\long\def\optional#1{\empty}

\def\bi{\begin{itemize}}

\def\ei{\end{itemize}}
\def\beq{\begin{equation}}
\def\eeq#1{\label{#1}\end{equation}}
\def\ba{\begin{array}}
\def\ea{\end{array}}

\def\sm{\hbox{\rm SM}}
\def\lpmln{\hbox{\rm LP}^{\rm{MLN}}}
\def\lpmln{{\rm LP}^{\rm{MLN}}}

\title{Explainable Fact Checking with Probabilistic Answer Set Programming}
\author{ 
Naser Ahmadi$^*$, Joohyung Lee$^\#$, Paolo Papotti$^*$, Mohammed Saeed$^*$\\
$^*$EURECOM, France ~~ $^\#$Arizona State University, USA \\
{\small \texttt{\{naser.ahmadi,papotti,mohammed.saeed\}@eurecom.fr, joolee@asu.edu}}}

\begin{document}

\maketitle

\begin{abstract}
One challenge in fact checking is the ability to improve the transparency of the decision. We present a fact checking method that uses reference information in knowledge graphs (KGs) to assess claims and explain its decisions. KGs contain a formal representation of knowledge with semantic descriptions of entities and their relationships. We exploit such rich semantics to produce interpretable explanations for the fact checking output. As information in a KG is inevitably incomplete, we rely on logical rule discovery and on Web text mining to gather the evidence to assess a given claim. Uncertain rules and facts are turned into logical programs and the checking task is modeled as an inference problem in a probabilistic extension of answer set programs. Experiments show that the probabilistic inference enables the efficient labeling of claims with interpretable explanations, and the quality of the results is higher than state of the art baselines. \end{abstract}

\section{Introduction} \label{sec:intro}
Due to the increase of sources spreading false information, computational {\em fact checking} has been proposed to support journalists and social media platforms with automatic verification of textual content~\cite{full_fact}. We focus on \textit{claims} that contain factual statements, such as ``William Durant was the founder of Chevrolet", and their verification against reference data, i.e., {\em Knowledge Graphs} (KGs). Assuming entities and relations involved in ``worth-checking'' claims have been identified~\cite{HassanALT17,JaradatGBMN18}, KGs are exploited to compute the {\em veracity} of claims expressed as structured data.

A KG is a structured representation of information which stores real-world entities as nodes, and relationships between them as edges. Entities and relations have semantic descriptions in the form of types and properties associated with them. 
KGs store large amounts of factual information and several of them are publicly available~\cite{suchanek2007yago}. 
For example, the English version of DBpedia stores 6M entities and 9B relation triples. 

Given a KG $K$ and a claim $f$, several approaches have been developed to estimate if $f$ is a valid claim in $K$.
In some of these methods, facts in the KG are leveraged to create features, such as paths~\cite{shi2016discriminative,ciampaglia15computational} or embeddings~\cite{bordes2013translating,socher13reasoning}, which are then used by classifiers to label as true or false a given test claim.  Other methods rely on searching for occurrences of the given claim on Web pages~\cite{Vault14,PopatMYW18}. However, such models are based on Machine Learning (ML) classifiers that in the best case can report the source of evidence for a decision but \textit{lack the ability to provide comprehensible descriptions} of how a decision has been taken for a given claim. 

\begin{table*}
\small
 \begin{lstlisting}[escapeinside={(*}{*)}]
0.75: foundedBy(a,b) (*$\leftarrow$*) keyPerson(a,b), foundedBy(c,b), product(c,d), product (a,d).
0.76: foundedBy(a,b) (*$\leftarrow$*) distributor(c,b),distributor(c,d),foundedBy(a,d).
0.97: negfoundedBy(a,b) (*$\leftarrow$*) foundingYear(a,c), birthYear(b,d), >(d,c).
0.56: negfoundedBy(a,b) (*$\leftarrow$*) foundedBy(a,c),relative(d,c),occupation(d,b).
0.67: negfoundedBy(a,b) (*$\leftarrow$*) parentCompany(b,c), subsidiary(c,d), parentCompany(d,a).
 \end{lstlisting}
 \caption{Example of discovered rules with their support for predicate {\em foundedBy} in DBpedia.}
    \label{tab:rules}
\end{table*}

To address this problem and effectively support transparent content moderation,
we use existing KGs as sources of evidence, together with logical reasoning to make fact checking decisions. The key idea is to assess as true or false a given claim and to provide \textit{human-interpretable explanations} for such decision in the form of supporting and contradicting evidence. Declarative Horn rules defined over the KG, such as those in Table~\ref{tab:rules}, guide the decision process and provide semantic arguments for the conclusion. For example, ``William Durant was the founder of Chevrolet" is marked as true and justified by the facts that Durant is a \textit{key person} for Chevrolet and he \textit{founded} another car company. This explanation comes from the first rule in the table\footnote{\textit{Product} is a predicate in the KG modeling the pairs (company \textbf{c}, product \textbf{p} of company \textbf{c}).}. 
On the other hand, ``Elon Musk was the founder of Chevrolet" is marked as false with the explanation that Musk was born after the company foundation year (third rule in the table). 

Unfortunately, two issues make the generation of such explanations hard. First, in general KGs do not come with the rich rules we need in our task. To address this issue, we exploit rule mining approaches \cite{galarraga2015fast,OrtonaMP18}, which automatically learn logical rules for a given KG (e.g., Table~\ref{tab:rules}). Second, KGs have data quality issues, due to the automatic methods that are used to build them at scale. Information stored in KGs is inevitably incomplete (Open World Assumption - OWA) and noisy, because of errors coming from the sources and the automatic extractors~\cite{Vault14}. For these reasons, in many cases, rules cannot be triggered. We identify these cases and resort to mining Web pages to get evidence for missing facts that are crucial to reach a decision for a claim~\cite{PopatMYW18}.

Discovering rules and mining facts from the Web enable a fully automatic system, but a new challenge arises from these approaches. In fact, both rules and mined facts are \textit{uncertain}, i.e., they come with a measure of the probability of being correct, which is some cases can be quite low. To address this third challenge, we use probabilistic answer set programming~\cite{lee2016weighted}. The reasoner is the key enabler of the inference that combines all the evidence in producing a fact checking decision with its explanation. 

Our main contribution is a fully \textit{automated and interpretable fact checking system} that effectively exploits uncertain evidence. Experimental results on several predicates on a real KG show that our method (i) obtains qualitative results that are comparable or better than existing black-box ML methods and (ii) outputs human consumable explanations.

We introduce the background on the existing systems and the problem definition in Section~\ref{sec:prem}. We then discuss our methods in Section~\ref{sec:framework} and experimental results in Section~\ref{sec:exp}. We conclude with a discussion of related work and open problems in Section~\ref{sec:rw} and Section~\ref{sec:conc}, respectively.

\section{Preliminaries} \label{sec:prem}
In this section, we describe the main building blocks of our framework and define our problem.

\vspace{1ex}
\noindent{\bf Knowledge Graph.} An RDF KG is a database representing information with triples (or \textit{facts}) 
$p(s,o)$ where a \emph{predicate} $p$ connects a \emph{subject} $s$ and an \emph{object} $o$. 
For example, the fact that E. Musk was born in 1971 is expressed with a triple {\em birthYear}(E. Musk, 1971). 
In a triple, the subject is an \emph{entity}, i.e., a real-world concept; the object is either an entity or a \emph{literal}, i.e.,  primitive types such as number, date, and string; and the triple predicate specifies a relationship between subject and object. We call a triple to be assessed a {\em claim}.

\vspace{1ex}
\noindent{\bf Rule Mining in KGs.}
In our framework, we exploit algorithms for mining declarative Horn rules from KGs~\cite{galarraga2015fast,OrtonaMP18}.
A Horn rule has the form:
\begin{equation} \label{eq:1}
h(x,y) \leftarrow B
\end{equation}
where $h(x, y)$ is a single atom (head of the rule) and $B$ (body of the rule) is a conjunction of atoms 
$$B_{1}(z_1, z_2) \wedge B_{2}(z_3, z_4)\wedge \dots \wedge B_{n}(z_{2n-1}, z_{2n}).$$
An atom is a predicate connecting two variables, two entities, an entity and a variable, or a variable and a constant (string or number). 
A mining algorithm outputs positive rules (e.g., \textit{spouse} in the head), which identify relationships between entities, e.g., ``if two persons have a child in common, they are in the spouse relation'', and negative rules (\textit{negspouse} in the head), which identify data contradictions, e.g., ``if two persons are in the parent relation, one cannot be the spouse of the other''. 

A fact is derived from a rule if all the variables in the body of the rule can be replaced with constants in the KG. For example, consider again Table~\ref{tab:rules} and the negative rule:
\textit{negfoundedBy(a,b)} $\leftarrow$ \textit{foundingYear(a,c), birthYear(b,d),} $>$\textit{(d,c)}.
We can derive \textit{negFoundedBy}(E. Musk, Chevrolet) because there is a replacement for the body of the rule, i.e, 
``foundingYear(Chevrolet,1911), birthYear(E. Musk,1971), $>$(1971,1911)''.



For rule mining, we adopt {\sc RuDik}~\cite{OrtonaMP18}.
For every predicate in the KG, the mining algorithm 
outputs rules together with a measure of support.

\vspace{1ex}\noindent{\bf Assessment of Claims on the Web.} As KGs are usually incomplete, we exploit also textual documents for our analysis. Text mining systems get as input a claim $c$ expressed in natural language and analyze $c$'s credibility w.r.t. relevant Web documents. The systems exploit the
joint interaction among language style of documents, their stance
towards a claim, and source trustworthiness.

For example, consider the claim \textit{foundedBy}(Chevrolet, W. Durant), which is not in the KG, and positive rule from Table~\ref{tab:rules}:
\textit{foundedBy(a,b)} $\leftarrow$ \textit{keyPerson(a,b), foundedBy(c,b), product(c,d), product(a,d)}.
Assume the KG contains the facts \textit{keyPerson}(Chevrolet, W. Durant), \textit{foundedBy}(GM, W. Durant), and \textit{product}(GM, Automobile), but it misses the product information for Chevrolet. Because of the OWA, we do not know if this is a false fact, or a true one missing from the KG. 
We therefore test \textit{product}(Chevrolet, Automobile) and the text mining system says that, according to Web documents, the fact is true with confidence 0.57.

In our framework, we adopt {\sc CredEye}, a state of the art system for the automatic credibility assessment of textual claims~\cite{Popat:2016}.
To extract Web articles relevant to the input claim, it uses a commercial search engine (i.e., Bing).
Each document 
is divided into a set of overlapping snippets, and snippets that are strongly related to the claim in terms of unigram and bigram 
are extracted. Snippets are then used to compute \textit{support} and \textit{refute} scores with logistic regression classifiers trained on claims and evidence documents from the Snopes fact checking repository. 
The scores are fed as features into a classifier with L1-regularization, distantly trained on Snopes. 

\vspace{1ex} \noindent{\bf Probabilistic Answer Set Programming.}
Given a claim, we collect the rules with their confidence, the evidence (from the KG and the Web sites), and cast fact checking as a reasoning problem. For this task, we adopt
$\lpmln$~\cite{lee2016weighted}, a probabilistic extension of answer set programs with the concept of weighted rules as in Markov Logic. 
In ASP, search problems are reduced to computing {\em stable models} (or answer sets), a set of beliefs that hold for the given problem.
In $\lpmln$, a weight is assigned to each rule so that the more rules a stable model satisfies, the larger weight it gets, and the probability of the stable model is computed by normalizing its weight among all stable models. In our setting, given a set of rules and evidence facts, we want to see if the given claim is in the stable model.

An $\lpmln$ program $\Pi$ is a finite set of weighted rules of  the form:
\beq
w: A \leftarrow B 
\eeq{lpmln-rule}
where $A$ is a disjunction of atoms, $B$ is a conjunction of literals (atoms and negated atoms), and $w$ is a real number or the symbol $\alpha$.
When $A$ is $\bot$ (the empty disjunction), the rule asserts that $B$ should be false in the stable model.
An $\lpmln$ rule \eqref{lpmln-rule} is called {\em soft} if $w$ is a real number or {\em hard} if $w$ is $\alpha$.
An $\lpmln$ program is {\em ground} if its rules contain no variables. 
Any $\lpmln$ program $\Pi$ of signature $\sigma$ with a ground $\lpmln$ program $gr_{\sigma}[\Pi]$ is obtained from the rules of $\Pi$ by replacing every variable with every ground term of $\sigma$ with constants from the evidence facts. The weight of a ground rule in $gr_{\sigma}[\Pi]$ is the same as the weight of the rule in $\Pi$ from which the ground rule is obtained. 
By $\overline{\Pi}$ we denote the unweighted logic program obtained from $\Pi$,  i.e., $\overline{\Pi} =\{R \hspace{0.1cm}|\hspace{0.1cm} w:R\in\Pi\}$. 

For an $\lpmln$ program $\Pi$, ${\Pi}_I$ denotes the set of rules $w: R$ in $\Pi$ such that $I\models R$ and {\rm SM}$[\Pi]$ denotes the set $\{I \mid \text{$I$ is a (deterministic) stable model of $\overline{\Pi_I}$}\}$. The (unnormalized) weight of $I$ under $\Pi$ is defined as:
\[
\begin{aligned}
    W_{\Pi}(I) = 
    \begin{cases}
        exp(\sum\limits_{w:R\in\Pi_I} w) \  \text{ if } I \in \sm[\Pi];\\
        0 \hspace{4.5px} \hspace{2cm} $otherwise$.
        \end{cases}
\end{aligned}
\]
The probability of $I$ under $\Pi$ is the normalized weight defined as:
$P\textsubscript{$\Pi$}(I) = \lim_{\alpha\to\infty} \frac{W\textsubscript{$\Pi$}(I)}{\sum_{J\in {\rm SM}[\Pi]} W\textsubscript{{$\Pi$}}(J)}.
$



{\sc lpmln2asp}~\cite{lee17computing} 
is an implementation of $\lpmln$ using ASP solver {\sc clingo}.  
The system returns the most probable stable models (answer sets). In our problem formulation, given a claim $p$(x,y), we identify all the rules that have predicate $p$ and {\em negp} in the conclusion and the evidence facts for the bodies of the rules. We then run {\sc lpmln2asp} and check if $p$ or {\em negp} are in the answer set.

\vspace{1ex}
\noindent{\bf Problem Statement.}
Given an input claim to be verified and a KG, our goal is to compute an assessment of the veracity of the claim 
and the explanations for such decision, expressed as the union of substitutions for the body of the rules that have triggered the inference in the reasoner. 

The uncertainty in the discovered rules and in the facts extracted from the Web make the problem challenging and the role of the reasoner important.

\vspace{1ex}
\noindent{\bf Limits of Existing Solutions.}
Both the text mining and the rule generation system can be used individually as fact checking tools according to our problem definition. However, they both have strong limitations. The uncertain rules alone cannot make a clear assessment decision in many cases  because of (i) conflicting rules both supporting and refusing a fact at the same time, and (ii) lack of evidence in the KG. The Web mining cannot provide semantic explanations and also suffers from the cases where there is no enough evidence to obtain an answer. These limitations apply also for other ML fact checking systems~\cite{ciampaglia15computational,shi2016discriminative,bordes2013translating,socher13reasoning} and motivate our choice to use a unified framework to combine both sources of signals with a probabilistic reasoner.

\begin{figure}[t]
		\includegraphics[width=1.04\columnwidth]{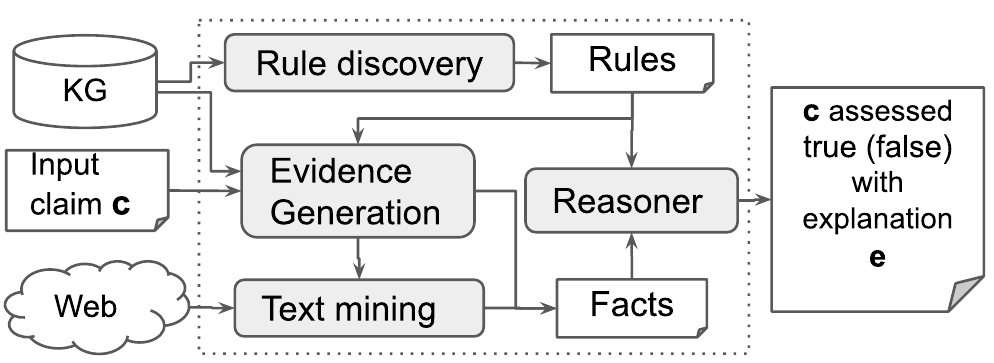}	
		\vspace{-1ex}
		\caption{Our Fact checking framework.}
	\label{fig:framework}
		\vspace{-0.5ex}
\end{figure}

\section{Framework} \label{sec:framework}

Figure~\ref{fig:framework} shows our framework. 
The \textit{Rule discovery} module 
takes as input the KG $K$ to generate the 
rules. We then convert the discovered rules $\Sigma$ into the input language of the reasoner, 
where the weight of a rule is its {\em support}.
For the given claim $c: p(x,y)$, $p \in K$ and rules $\Sigma$, the \textit{Evidence Generation} module collects relevant evidence facts (triples satisfying the body of the rules) from the KG and from Web with the \textit{Text mining} module. 
We then feed rules and evidence to the 
\textit{Reasoner} module, where different modes of computation can be used to infer if $p(x,y)$ or $negp(x,y)$ is in the answer set. The reasoner output includes a human-interpretable explanation for the decision. The details of the main steps are given next. 




\subsection{Rule Generation}
\label{sec:ruleGen}
Consider a claim $c: p(x,y)$ with $p \in K$, our first step is to obtain the set of rules $\Sigma$. 

\vspace{1ex}
\noindent
{\bf Rule Discovery}: The rule discovery module starts by generating $M$ positive and $M$ negative examples for $p$. Positive examples are $(x,y)$ entity pairs s.t. $p(x,y)\in K$, and negative examples are $(x,y)$ pairs that satisfy the following conditions~\cite{OrtonaMP18}: 
\begin{itemize}
  \setlength{\itemsep}{1.2pt}
  \setlength{\parskip}{0pt}
  \setlength{\parsep}{0pt}
  \item $p(x,y) \notin K$;
  \item there is either some $y' \ne y$ s.t. $p(x,y') \in K$ or some $x'\ne x$ s.t. $p(x',y) \in K$;
  \item there is some $p' \ne p$ s.t. $p'(x,y) \in K$.
\end{itemize}

{\sc RuDik} uses the examples and the KG to mine positive and negative rules ($\Sigma$) for $p$.

Consider the mining of positive rules for predicate {\it spouse}. Positive examples are pairs of married people and negative examples are pairs of people who are not married to each other. 
Given the examples, the algorithms output approximate rules, i.e., rules that do not necessarily hold over all the examples, as those are derived from a noisy and incomplete KG. 
The example sets switch role for the discovery of negative rules, i.e., not married people play the role of the positive examples.

As in association rule mining, the support $s$ of each rule is computed as the {\em support value} of the rule divided by the number of examples used in the rule discovery step~\cite{galarraga2015fast}. 


\vspace{1ex}
\noindent{\bf Convert Rules into $\lpmln$}: Rules in $\Sigma$ are rewritten into the input language of {\sc lpmln2asp} with their weights.
For instance, for the {\em spouse} predicate, a positive rule 
is rewritten into $\lpmln$ as
\beq
w:\ spouse(a,b) \leftarrow child(a,c), parent(c,b).
\eeq{spouserule}

An original support $s$ equals to 0 corresponds to a weight $w$ of $-\infty$ and a support of 1 to a weight of $+\infty$. We convert the rule support into a weight for a program with the equation:
 $ w= \ln{\frac{s}{1-s}}$.




 


\vspace{1ex}
\noindent{\bf Generic Rules}:
We add two rules to the set associated to each predicate. These rules are generic and model natural constraints that play an important role in our fact checking system.

The first rule ensures that {\em p(x,y)} and {\em negp(x,y)} cannot be true at the same time, i.e., a claim should not be assessed as false and true. This is a hard rule, which is always valid. 
\beq
  \alpha:\ \   \bot \leftarrow p(x,y), negp(x,y)
\eeq{constr1}

The second rule enforces the \textit{functionality} of a predicate. If a predicate is functional, such as the predicate expressing the capital of a country, then there is only one value that can be in the solution. However, this is not true for all predicates, e.g., a person can be the author of several books. The support of the rule models the functionality of the predicate. 
We express this constraint stating that a claim cannot have two different object values. 
\beq
   w:\ \ \bot \leftarrow p(x,y), p(x,z), y \neq z
\eeq{constr2}

These generic rules steer the reasoner in the computation of the truthfulness/falseness probability for the input claim.

\subsection{Evidence Generation} 
For a given claim, we execute the following steps to gather the evidence for a fact checking decision.

\vspace{1ex}
\noindent{\bf Generate Evidence Triples from KG:} 
For each rule in $\Sigma$, we substitute the head variables with the values of the claim and collect all the triples in the KG that have a valid substitution to its body. 
More precisely, the head variables in the body of a rule are constrained to the value of the subject and object of the claim. 
Then, the {\em evidence triples} are identified by querying the KG with the rewritten body of the rule.
For example, given the \textit{spouse} rule above and claim \textit{spouse}(Mike,Laure), the body is rewritten as a query: {\small {\tt child(Mike,\textit{c}), parent(\textit{c},Laure)}}, where \textit{c} is a universal variable.


\vspace{1ex}
\noindent{\bf Generate Evidence Triples from Web:} 
Our reasoner models also the uncertainty for the evidence facts. The KG is considered trustworthy, so the weights for the evidence from the KG are set at infinite. However, because of the OWA, we cannot find every true fact in the KG. For claims for which no rule can be executed, we resort to a Web text mining system~\cite{Popat:2016}. 
For each rule, we substitute the subject and the object according to the input claim. 
If a single atom is non-replaceable with KG facts in the body of a rule, 
then we use the Web module to validate the missing fact. Notice that only grounded facts can be verified with the Web module, such as \textit{child}(Mike,Marie).
If the rewritten body contains a fact with a variable, such as \textit{child}(Mike,\textit{c}) above, we discard the claim. If the Web module returns a probability $p$ of a fact being correct greater than 0.5, than we add it to our evidence. 

As an example, consider the positive rule: \textit{locatedIn(x,y)} $\leftarrow$ \textit{hasCapital(z,x), locatedIn(x,y)}, the claim \textit{locatedIn}(Sacramento, USA), and a KG with fact \textit{hasCapital}(CA, Sacramento). Assuming that the fact for CA located in USA is missing from the KG, we query the Web module for \textit{locatedIn}(CA, USA). 

Similarly to the conversion of the rule support into the weight of an {LP\textsuperscript{MLN}} program (Section~\ref{sec:ruleGen}), we convert the probability $p$ of a fact of being true into a weight $w$ for the fact when we use it as evidence for the reasoner. 

\BOCC
\item {\bf Inference using {LP\textsuperscript{MLN}}}: For each query triple in test data, we use $\lpmln$ to infer the likelihood of the relation between entity pair.
\EOCC

\subsection{Inference for Fact Checking}
We discuss two inference methods that enable us to expose the rules and the evidence triples involved in a decision for a claim $p(x,y)$.

\begin{itemize}
  \setlength{\itemsep}{1.2pt}
  \setlength{\parskip}{0pt}
  \setlength{\parsep}{0pt}
\item {\bf Pure ASP} checks if $p(x,y)$ or $negp(x,y)$ is in the stable model of the rules without including the rule weights. This method only states if the positive or negative triple for the claim can be derived. Since we rely on Horn rules, there is only one stable model for them. If the stable model contains both $p(x,y)$ and $negp(x,y)$, it violates  constraint~\eqref{constr1}, so we conclude neither $p(x,y)$ nor $negp(x,y)$. A similar case happens when the stable model violates the functionality of a predicate.
  
\item {\bf $\lpmln$ MAP inference with weighted rules} checks if $p(x,y)$ or $negp(x,y)$ is in the most probable stable model of the weighted rules using {\sc lpmln2asp}. This method utilizes the weighted rules and the evidence facts to find a more likely answer at the cost of violating constraints \eqref{constr1} and \eqref{constr2}. 
  

\end{itemize}




\lstset{
   frame=single,
   basicstyle=\fontsize{8}{10}\selectfont\ttfamily,
   numbers=left,
   numbersep=4pt,
    breaklines=true,
   numberstyle=\tiny,  
   stringstyle=\small\ttfamily,
   showspaces=false,
   showstringspaces=false,
   literate={~} {$\sim$}{1}
}

\vspace{1ex}
\noindent {\bf Example.} We want to check if {\em Glen Cook} is the author of the book {\em Cold Copper Tears}.
The following weighted rules are mined from the KG\footnote{For readability, we report normalized support (confidence) for rules (evidence triples), instead of weights.}:

\begin{lstlisting}[escapeinside={(*}{*)}]
0.04: author(A,B) (*$\leftarrow$*) runtime(A,C), activeYearsStartYear(B,D), C<D.
0.04: author(A,B) (*$\leftarrow$*) birthYear(B,C),    runtime(A,D), C>D.
0.13: author(A,B) (*$\leftarrow$*) author(C,B), subsequentWork(A,C).
0.02: author(A,B) (*$\leftarrow$*) previousWork(A,C), literaryGenre(C,D),genre(B,D).
0.02: negauthor(A,B) (*$\leftarrow$*) writer(C,B),      format(C,D), format(A,D).
0.38: negauthor(A,B) (*$\leftarrow$*) runtime(A,C), activeYearsStartYear(B,D), C<D.
0.31: negauthor(A,B) (*$\leftarrow$*) birthYear(B,C), runtime(A,D), C>D.
0.02: negauthor(A,B) (*$\leftarrow$*) writer(C,B), previousWork(C,A).
0.02: negauthor(A,B) (*$\leftarrow$*) writer(C,B), previousWork(C,D), subsequentWork(A,D).
0.08: negauthor(A,B) (*$\leftarrow$*) writer(C,B),     genre(C,D), genre(A,D).
0.02: negauthor(A,B) (*$\leftarrow$*) writer(C,B), subsequentWork(C,A).
0.02: negauthor(A,B) (*$\leftarrow$*) previousWork(A,C), subsequentWork(D,C), writer(D,B).
   (*$\alpha:$*) (*$\bot \leftarrow$*) negauthor(A,B), author(A,B).
0.04: (*$\bot \leftarrow$*) author(A,B), author(A,C), B(*$\neq$*)C.
\end{lstlisting}

Notice that not all rules are semantically correct: rule 1 is not valid (and has low support), while rule 3 is correct in most cases (in fact it has a higher support). Notice also rule 13, which is the hard constraint stating that a fact cannot be true and false at the same time and rule 14 reflecting the low functionality for the \textit{author} predicate.
The evidence generator module collects the following triples from the KG (facts with confidence 1) and the Web mining module (all other facts):

\begin{lstlisting}[escapechar=!,numbers=none]
0.55: literaryGenre('Cold_Copper_Tears','Fantasy').
0.52: literaryGenre('Cold_Copper_Tears','Mystery_fiction').
1: !\bf{previousWork('Cold Copper Tears','Bitter Gold Hearts').}!
0.69: !\bf{subsequentWork('Cold Copper Tears','Old Tin Sorrows').}!
0.56: activeYearsStartYear('Glen_Cook','1970').
0.59: author('Bitter_Gold_Hearts','Glen_Cook').
1: !\bf{author('Old Tin Sorrows','Glen Cook').}!
1: !\bf{genre('Glen Cook','Fantasy').}!
1: genre('Glen_Cook','Science_fiction').
1: literaryGenre('Bitter_Gold_Hearts','Mystery_fiction').
1: !\bf{literaryGenre('Bitter Gold Hearts','Fantasy').}!
1: literaryGenre('Old_Tin_Sorrows','Mystery_fiction').
1: literaryGenre('Old_Tin_Sorrows','Fantasy').
1: previousWork('Bitter_Gold_Hearts','Sweet_Silver_Blues').
1: previousWork('Old_Tin_Sorrows','Cold_Copper_Tears').
1: subsequentWork('Bitter_Gold_Heart','Cold_Copper_Tears').
1: subsequentWork('Old_Tin_Sorrows','Dread_Brass_Shadows').
1: author('The_Black_Company','Glen_Cook').
1: genre('The_Black_Company','Dark_fantasy').
1: genre('The_Black_Company','Epic_fantasy').
1: genre('The_Black_Company','Fantasy_novel').
\end{lstlisting}

The $\lpmln$ inference outputs that the input fact is true because of rules 3 and 4 together with the facts in bold in the evidence set. Here, {\em Old Tin Sorrows} is the {\em subsequentWork} of {\em Cold Copper Tears} whose author is {\em Glen Cook}. These two facts satisfy the body of rule 3 to derive the {\em author} relation between {\em Cold Copper Tears} and {\em Glen Cook}. Similarly, for rule 4, {\em Fantasy} is the {\em genre} of {\em Glen Cook}, which is also the {\em literaryGenre} of book {\em Bitter Gold Hearts}. Further, {\em Bitter Gold Hearts} is the {\em previousWork} of {\em Cold Copper Tears}. This sequence of three facts in the evidence set satisfies the body of rule 4 to derive the {\em author} relation between the test entities.
By using the MAP inference, 
we can find in the answer set:
\begin{lstlisting}[numbers=none]
author(Cold_Copper_Tears,Glen_Cook)
\end{lstlisting}


\section{Experiments} \label{sec:exp}
We test our proposal against baseline methods over claims from a real KG. 

\begin{table}[h]
    \centering
   {\small
    \begin{tabular}{|c|c|c|c|c|} 
    \hline
        &\textit{spouse} & \textit{deathPl.}& \textit{vicePres.}& \textit{almaMater} \\
        \hline
        Positive & 22 & 25 & 65 & 27 \\
        \hline
        Negative & 72 & 33 & 27 & 21 \\
        \hline
    \end{tabular}}
    \caption{Number of discovered rules for each predicate.}
        \vspace{-1ex}
    \label{tab:ruleStats}
\end{table}

\vspace{1ex}\noindent {\bf Datasets.}
From the latest (online) DBpedia, we selected four predicates $P$ = \textit{spouse}, \textit{deathPlace}, \textit{vicePresident}, \textit{almaMater}.
In the following, all rules have been mined from 2K positive and 2K negative examples. Statistics for the discovered rules are reported in Table~\ref{tab:ruleStats}.

We create 3 datasets with each containing 100 true and 100 false facts for every predicate, for a total of 2400 claims. 
True facts are randomly taken from the KG, false ones are created according to the procedure described in the previous section. 
True facts are then removed from the graph.

\vspace{1ex}\noindent {\bf Metrics.}
In the output, we evaluate the results by distinguishing the following cases. For each test fact, we count \textit{correctly} labelled facts (T) and 
\textit{incorrectly} labelled ones (F). 
We also count {\em Undecided} (U), when the method at hand cannot make a decision\footnote{In the reasoner, neither {\em p(x,y)} or {\em negp(x,y)} are in the stable model.}.
We also use precision, defined as {\small (T)/(T+F)}, recall, defined as {\small (T)/(T+F+U)}, and their combination in the F-score (harmonic mean).

\vspace{1ex}\noindent {\bf Methods.}
We run three baseline methods.
The first is the Web text miner {\sc CredE}~\cite{Popat:2016}. 
The second is the state of the art link prediction method for KGs {\sc Kgm}~\cite{shi2016discriminative}, which uses the graph facts as training data and a ML classifier. The third baseline is the application of the discovered rules, without considering their weights ({\sc ASP}), i.e.,  $\lpmln$ MAP inference with hard rules. The first two approaches ({\sc CredE} and {\sc Kgm}) cannot provide explanations, while the third ({\sc ASP}) does not exploit the reasoning. We identified 0.5 as the threshold value for both {\sc CredE} and {\sc Kgm} to maximize their F-score.

We consider two variants of our solution. The first is the $\lpmln$ MAP inference with weighted rules over the KG data only ({\sc MAP}). The second is {\sc MAP} integrated with evidence collected from the KG and Web documents ({\sc MAP+W}). 
For these methods, we check if the claim is in the stable model, i.e., it can be inferred. 

\begin{table}[h]
    \footnotesize
    \begin{tabular}{|c|c|c|c|c|c|} 
        \hline
          & \textit{almaMat.}  & \textit{deathPl.} & \textit{spouse} &  \textit{vicePres.} \\
        \hline
         {\sc CredE} & .41(.03) & .59(.06) & .44(.07) & .36(.15) \\
        \hline
         \
         {\sc Kgm} & .73(.08) & .68(.01) & .86(.01) & \textbf{.81}(.03)\\
        \hline
         \
         {\sc ASP} & .70(.06) &.01(.01) & .31(.08) & .18(.16)\\
        \hline
        \
         {\sc MAP} & \textbf{.88}(.14) &.75(.15) & \textbf{.87}(.11) & .66(.22)\\
         \hline
         \
         {\sc MAP+W} & .\textbf{88}(.09) &.\textbf{83}(.11) & .86(.10) & {.68}(.18)\\
         \hline
    \end{tabular}
    \vspace{-1ex}
\caption{Average F-score results (SD) for four predicates with all methods over 3 datasets.}
\label{tab:full_fscore}
\end{table}

\vspace{1ex}\noindent {\bf Results.}
Table~\ref{tab:full_fscore} reports F-score results and standard deviation (SD) for true and false claims averaged over the 3 datasets. For two predicates, {\sc MAP+W} is the best method in terms of F-score, with an average over all predicates of 0.81, followed by {\sc MAP} with .79 and {\sc Kgm} with .77. For all predicates 
method {\sc ASP} has very poor performance because of a large number of claims with no rule to be grounded with the KG evidence. Several of these claims are solved with the reasoner in {\sc MAP} with high precision (1 in most cases) but not perfect recall. Web evidence in {\sc MAP+W} also enables the triggering of more rules, but at the cost of a lower precision because the text miner is not always reliable, as it is clear from the results for {\sc CredE}. 

The issue of undecided claims affects heavily the results for predicate \textit{vicePresident} in all methods based on rules. 
In general, there is no clear correlation between the number of rules and the quality of the results for the rule-based methods. This suggests that the quality of the rules (and of the evidence facts) is more important than their number. Also, more functional predicates, such as \textit{spouse}, are easier to fact check for most methods~\cite{HuynhP18}.

\begin{table*}[t]
    \footnotesize
    \begin{tabular}{|c||c|c|c|c|c||c|c|c|c|c|} 
        \hline
         &  \multicolumn{5}{c||}{True claims}  &  \multicolumn{5}{c|}{False claims}  \\
        \hline
         & {\sc CredE} & {\sc Kgm}& {\sc ASP}& {\sc MAP}& {\sc MAP+W} &{\sc CredE} & {\sc Kgm}&  {\sc ASP}& {\sc MAP}& {\sc MAP+W}\\
        \hline
        Correct(/100) & 50(8)  & 96(1) & 1(1) & 58(27)  & 62(31)  & 23(8) & 7(2) &  0 & 62(14) & 78(8)\\
        \hline
        Incorrect(/100) & 19(3) & 4(1) &  0 & 0  & 21(18)  & 55(1) & 93(2) & 0 & 11(4) & 5(4)\\
         \hline
        Undecided(/100) & 31(9) & 0 & 99(1) & 42(27)  & 17(13)  & 22(7) & 0 &  100(1) & 28(18) & 17(9)\\
        \hline \hline
        Precision & .72 & .96 & 1 & 1 & .75
        & .29 & .07 & 1 & .85 & .94\\
        \hline
        Recall & .69 & 1 & .01 & .58 & .83
        & .78 & 1 & 0 & .72 & .83\\
        \hline
        F-score & .70 & .98 & .01 & .74 & .79
        & .43  & .13 & .01 & .78 & .88\\
        \hline
    \end{tabular}
    \vspace{-1ex}
\caption{Average results (SD) for {\em deathPlace} predicate with all methods over 3 datasets.}
\label{tab:full_spouse}
\end{table*}

Table~\ref{tab:full_spouse} reports a detailed analysis for predicate \textit{deathPlace} with average results over the 3 datasets. 
The first evidence is that {\sc Kgm} has the best performance for true claims but falls behind {\sc MAP} methods for false ones.
Neither {\sc CredE}
performs well with false claims. We emphasize that in fact checking false claims are more important.

Results for the rule-based methods show that reasoning is key for our approach.
For true claims, {\sc ASP} correctly labels only 1\% of the test facts, while the {\sc MAP} labels 58\% of them without mistakes on average. 
{\sc ASP} suffers the facts for which there is a contradiction among the positive and the negative rules, while {\sc MAP} inference makes the right decision by exploiting the weights of the rules.
However, for 42 true claims on average, none of the rules are triggered in {\sc MAP}. The coverage is increased by adding more evidence with the Web mining module ({\sc MAP+W}), at the cost of a lower precision but better overall F-score. 
The benefit of rules and Web evidence is clearer with false claims. While in this setting 
{\sc CredE} and {\sc Kgm} show poor results, {\sc MAP+W} reports high precision (94\% on average) and an average recall of 83\%, with a very significant increase in all metrics compared to {\sc MAP}. From a manual verification, we explain the better results for false claims with 
the better quality of the negative rules w.r.t. positive ones for \textit{deathPlace}, i.e., it is easier to find a negative rule than a positive rule for this predicate. This is consistent with previous rule quality assessments~\cite{OrtonaMP18}.

\begin{table}[t]
\small
 \begin{lstlisting}[escapeinside={(*}{*)},numbers=none]
 FALSE : almaMater(Michael White, UT Austin)
  (*$\leftarrow$*)  employer(Michael White, UT Austin)
  (*$\leftarrow$*)  occupation(Michael White, UT Austin)
  (*$\leftarrow$*)  almaMater(Michael White, Abilene Christian Univ.), almaMater(Michael White, Yale Divinity School)
\end{lstlisting}
 \caption{Example of {\sc MAP+W} output for claim {\em almaMater}(Michael White, UT Austin).}
    \label{tab:explain}
\end{table}

In all the cases for which an answer is produced, rule-based methods explain their decision by showing involved rules and corresponding evidence sets. This makes it relatively easy to identify what is the cause for a conclusion,
as for the example reported in Table~\ref{tab:explain}. The given claim is labeled as false because of the three rules that apply with evidence coming both from the KG and the Web.

\begin{table}[h]
    \footnotesize
    \begin{tabular}{|c|c|c|c|c|} 
        \hline
         & \textit{spouse}  &  \textit{deathPl.} & \textit{vicePres.} & \textit{almaMat.}  \\
        \hline
         {\sc CredE} &   6435     &  7377  &   7210     & 7355\\
         \hline
         {\sc Kgm} & 16 &  15   &   12  & 13 \\
         \hline
         {\sc ASP} & 7        &    8      &    9    & 8\\
         \hline
         {\sc MAP} &  475       &     822      &  1880  & 408 \\
         \hline
         {\sc MAP+W} & 485   &     1897  &  3448 & 409 \\
        \hline
    \end{tabular}
    \vspace{-1ex}
\caption{Average execution times (secs) for 200 claims.}
\label{tab:execTime}
\end{table}

Finally, we report on the execution times in Table~\ref{tab:execTime}. Methods {\sc Kgm} and {\sc ASP} are the fastest, with a single claim checked in less than 0.1 seconds. Despite we are not counting the time to gather Web pages, {\sc CredE} is the slowest method, with up to 37 seconds on average to check a claim. {\sc MAP} and {\sc MAP+W} are in the middle, taking from 2 to 17 seconds to check a claim on average. The time differences depend on the predicate at hand, as checking predicates with less evidence in KG requires more calls to the text mining module.

\section{Related Work} \label{sec:rw}
There are two main tasks in computational fact checking: (1) monitor and spot claims~\cite{HassanALT17,JaradatGBMN18}, (2) check claims and explain outcomes. We focus on the second task and on factual facts, specifically.
Related approaches try to align the fact to trusted data resources, such as KGs~\cite{ShiralkarFMC17}, Web documents~\cite{LehmannGMN12}, and databases~\cite{CaoMT18,wu2014toward}. 
These approaches create features for binary classifiers from the data in the KG. Features exploit the structure of the training examples, in the form of paths~\cite{shi2016discriminative,ciampaglia15computational} or geometric properties in a multi-dimensional space with embeddings~\cite{bordes2013translating,socher13reasoning}. As providing interpretable descriptions of the outcome of a ML model, such as SVM, is an active topic of research~\cite{nips2019interpretml}, we argue that semantically rich rules and their evidence facts are useful explanations for a fact checking outcome.

Markov Logic combines first-order logic and Markov networks \cite{richardson06markov}. In principle, learning in Markov Logic could learn the uncertain rules and inference can be applied to the learned rules as we do here. We tested {\em alchemy} to learn logical rules for {\em spouse} relation with only 10 positive examples, the system was not able to produce results after 2 hours of execution. This illustrates that rule learning in Markov Logic has scalability issues with large KGs such as DBpedia, let alone the quality of the rules learned.

ILP systems for rule discovery, such as {\sc ALEPH}~\cite{srinivasan2001aleph}, assume the closed world assumption and the input 
examples to be error-free.
These assumptions do not hold in KGs and 
{\sc RuDiK} outperform this kind of systems~\cite{OrtonaMP18}. Recently, other proposals have studied the problem of explainable fact checking with rules, but they focus on manually crafted constraints~\cite{Gad-Elrab0UW19,Leblay17}, while our system relies on discovered rules only. Experimental results on the same DBpedia predicates reported in previous work~\cite{Gad-Elrab0UW19} show that our solution performs better despite being fully automatic.


\section{Conclusion} \label{sec:conc}
We presented a fully automated fact checking framework based on KGs and Web documents as reference information. Given a fact expressed as a triple over entities in the KG, our method validates its veracity, with better average accuracy than state of the art ML methods, and provides an explanation of the decision by exposing facts that support or contradict the given claim according to a set of rules. The system does not rely on a human configuration, as rules are automatically discovered and additional information to complement the KG is mined from the Web.

An interesting direction for extending the framework is to include a module for claim detection and explore the opportunities of an end-to-end system~\cite{Thorne18Fever}.
A second direction is to exploit the information from the reasoner to steer the quality management of the KG~\cite{Vault14}. e.g., inspect undecided claims to identify parts of the KG that 
need data curation. Finally, we aim at integrating natural language generation techniques to produce explanations that are easier to read for the target users~\cite{gatt2018survey}.


\bibliographystyle{abbrv}
\bibliography{FC-bib} 

\begin{thebibliography}{10}

\bibitem{full_fact}
M.~Babakar and W.~Moy.
\newblock The state of automated factchecking.
\newblock \url{https://fullfact.org/blog/2016/aug/automated-factchecking/},
  2016.

\bibitem{bordes2013translating}
A.~Bordes, N.~Usunier, A.~Garcia-Duran, J.~Weston, and O.~Yakhnenko.
\newblock Translating embeddings for modeling multi-relational data.
\newblock In {\em NIPS}, pages 2787--2795, 2013.

\bibitem{CaoMT18}
T.~D. Cao, I.~Manolescu, and X.~Tannier.
\newblock Searching for truth in a database of statistics.
\newblock In {\em {WebDB}}, pages 4:1--4:6, 2018.

\bibitem{ciampaglia15computational}
G.~L. Ciampaglia, P.~Shiralkar, L.~M. Rocha, J.~Bollen, F.~Menczer, and
  A.~Flammini.
\newblock Computational fact checking from knowledge networks.
\newblock {\em PloS one}, 10(6), 2015.

\bibitem{Vault14}
X.~Dong, E.~Gabrilovich, G.~Heitz, W.~Horn, N.~Lao, K.~Murphy, T.~Strohmann,
  S.~Sun, and W.~Zhang.
\newblock Knowledge vault: a web-scale approach to probabilistic knowledge
  fusion.
\newblock In {\em {KDD}}, pages 601--610, 2014.

\bibitem{Gad-Elrab0UW19}
M.~H. Gad{-}Elrab, D.~Stepanova, J.~Urbani, and G.~Weikum.
\newblock Exfakt: {A} framework for explaining facts over knowledge graphs and
  text.
\newblock In {\em {WSDM}}, pages 87--95, 2019.

\bibitem{galarraga2015fast}
L.~Gal{\'a}rraga, C.~Teflioudi, K.~Hose, and F.~M. Suchanek.
\newblock Fast rule mining in ontological knowledge bases with {AMIE+}.
\newblock {\em The VLDB Journal}, 24(6):707--730, 2015.

\bibitem{gatt2018survey}
A.~Gatt and E.~Krahmer.
\newblock Survey of the state of the art in natural language generation: Core
  tasks, applications and evaluation.
\newblock {\em Journal of Artificial Intelligence Research}, 61:65--170, 2018.

\bibitem{HassanALT17}
N.~Hassan, F.~Arslan, C.~Li, and M.~Tremayne.
\newblock Toward automated fact-checking: Detecting check-worthy factual claims
  by claimbuster.
\newblock In {\em {KDD}}, 2017.

\bibitem{HuynhP18}
V.~Huynh and P.~Papotti.
\newblock Towards a benchmark for fact checking with knowledge bases.
\newblock In {\em Companion of TheWebConf ({WWW})}, pages 1595--1598, 2018.

\bibitem{JaradatGBMN18}
I.~Jaradat, P.~Gencheva, A.~Barr{\'{o}}n{-}Cede{\~{n}}o, L.~M{\`{a}}rquez, and
  P.~Nakov.
\newblock {ClaimRank}: Detecting check-worthy claims in {A}rabic and {E}nglish.
\newblock In {\em {NAACL-HTL}}, pages 26--30, 2018.

\bibitem{Leblay17}
J.~Leblay.
\newblock A declarative approach to data-driven fact checking.
\newblock In {\em {AAAI}}, pages 147--153, 2017.

\bibitem{lee17computing}
J.~Lee, S.~Talsania, and Y.~Wang.
\newblock Computing {L}{P}{M}{L}{N} using {A}{S}{P} and {M}{L}{N} solvers.
\newblock {\em Theory and Practice of Logic Programming}, 2017.

\bibitem{lee2016weighted}
J.~Lee and Y.~Wang.
\newblock Weighted rules under the stable model semantics.
\newblock In {\em KR}, pages 145--154, 2016.

\bibitem{LehmannGMN12}
J.~Lehmann, D.~Gerber, M.~Morsey, and A.~N. Ngomo.
\newblock Defacto - deep fact validation.
\newblock In {\em {ISWC}}, pages 312--327, 2012.

\bibitem{OrtonaMP18}
S.~Ortona, V.~V. Meduri, and P.~Papotti.
\newblock Robust discovery of positive and negative rules in knowledge bases.
\newblock In {\em {ICDE}}, pages 1168--1179, 2018.

\bibitem{Popat:2016}
K.~Popat, S.~Mukherjee, J.~Str\"{o}tgen, and G.~Weikum.
\newblock Credibility assessment of textual claims on the web.
\newblock In {\em {CIKM}}, 2016.

\bibitem{PopatMYW18}
K.~Popat, S.~Mukherjee, A.~Yates, and G.~Weikum.
\newblock Declare: Debunking fake news and false claims using evidence-aware
  deep learning.
\newblock In {\em EMNLP}, pages 22--32, 2018.

\bibitem{richardson06markov}
M.~Richardson and P.~Domingos.
\newblock Markov logic networks.
\newblock {\em Machine learning}, 62(1-2):107--136, 2006.

\bibitem{shi2016discriminative}
B.~Shi and T.~Weninger.
\newblock Discriminative predicate path mining for fact checking in knowledge
  graphs.
\newblock {\em Knowledge-Based Systems}, 104:123--133, 2016.

\bibitem{ShiralkarFMC17}
P.~Shiralkar, A.~Flammini, F.~Menczer, and G.~L. Ciampaglia.
\newblock Finding streams in knowledge graphs to support fact checking.
\newblock In {\em {ICDM}}, pages 859--864, 2017.

\bibitem{socher13reasoning}
R.~Socher, D.~Chen, C.~D. Manning, and A.~Ng.
\newblock Reasoning with neural tensor networks for knowledge base completion.
\newblock In {\em NIPS}, pages 926--934, 2013.

\bibitem{srinivasan2001aleph}
A.~Srinivasan.
\newblock The {Aleph} manual, 2001.

\bibitem{suchanek2007yago}
F.~M. Suchanek, G.~Kasneci, and G.~Weikum.
\newblock Yago: a core of semantic knowledge.
\newblock In {\em WWW}, pages 697--706. ACM, 2007.

\bibitem{Thorne18Fever}
J.~Thorne, A.~Vlachos, C.~Christodoulopoulos, and A.~Mittal.
\newblock {FEVER}: a large-scale dataset for fact extraction and verification.
\newblock In {\em NAACL-HLT}, 2018.

\bibitem{nips2019interpretml}
Workshop.
\newblock {Fairness and Explainability: From ideation to implementation}.
\newblock In {\em {NeurIPS}}, 2019.

\bibitem{wu2014toward}
Y.~Wu, P.~K. Agarwal, C.~Li, J.~Yang, and C.~Yu.
\newblock Toward computational fact-checking.
\newblock {\em Proceedings of the VLDB Endowment}, 7(7):589--600, 2014.

\end{thebibliography}

\end{document}